\documentclass[aps,11pt,tightenlines,nofootinbib]{revtex4}
\usepackage{amsmath,amscd,amssymb}
\usepackage{epsfig}

\def\beq{\begin{equation}}
\def\eeq{\end{equation}}
\def\be{\begin{eqnarray}}
\def\ee{\end{eqnarray}}
\newcommand{\dslash}{\partial \hskip -0.6em /}

\newcommand{\fract}[2]{\mbox{\small $\frac{#1}{#2}$}}

\newcommand{\zr}[1]{\mbox{\hspace*{#1em}}}
\newcommand{\ID}{\mbox{{\sf 1}\zr{-0.16}\rule{0.04em}{1.55ex}\zr{0.1}}}

\begin{document}

\title{Vacuum Energies of Non--Abelian 
String--Configurations in 3+1 Dimensions}

\author{H. Weigel$^{a)}$, M. Quandt$^{b)}$,
N. Graham$^{c)}$, O. Schr\"oder$^{d)}$}

\affiliation{
$^{a)}$Physics Department, Stellenbosch University,
Matieland 7602, South Africa\\
$^{b)}$Institute for Theoretical Physics, T\"ubingen University,
D--72076 T\"ubingen, Germany\\
$^{c)}$Department of Physics, Middlebury College,
Middlebury, VT 05753, USA\\
$^{d)}$science+computing ag, Hagellocher Weg 73, D--72070
T\"ubingen, Germany}

\begin{abstract}
We develop a method to compute the fermion contribution to the 
vacuum polarization energy of string--like configurations in a
non--abelian gauge theory. This calculation has been hampered 
previously by a number of technical obstacles. We use gauge 
invariance of the energy and separation of length scales in 
the energy density to overcome these obstacles. We present a 
proof--of--principle investigation that shows that
this energy is small in the $\overline{\rm MS}$
renormalization scheme. The generalization to other schemes is
straightforward.
\end{abstract}

\maketitle

\section{Introduction}

Various field theories suggest the existence of string--like 
configurations, which are the particle physics
analogues of vortices or  magnetic flux tubes in condensed
matter physics. Often they are called cosmic strings to distinguish
them from the fundamental variables in string theory and to indicate
that they stretch over cosmic length scales. If they exist, 
cosmic strings, can potentially have significant 
cosmological effects. We refer to ref.~\cite{Copeland:2009ga}
for a recent review on the physical implications of strings in the
standard model and beyond.

In the standard model of particle physics,
string solutions \cite{Vachaspati:1992fi,Achucarro:1999it,Nambu:1977ag}
are not topologically stable and thus can only be stabilized
dynamically. In exploring the existence of cosmic strings, it is 
therefore important to be able to accurately calculate their energies. 
Here we will develop and apply a method to compute the fermion contribution 
to the leading order quantum correction to the energy, the so--called
vacuum polarization or Casimir energy. In a large $N_C$ scenario with 
many internal degrees of freedom, the fermion contribution
dominates that of the bosons.  

A number of previous studies have investigated quantum properties of 
string and vortex configurations, invoking either approximations,
simplified configurations, or lower dimensions to cope with technical 
difficulties. Naculich~\cite{Naculich:1995cb} has shown that in the 
limit of weak coupling, fermion fluctuations destabilize the non--abelian 
Z--string. The quantum properties of $Z$--strings have also been
connected to non--perturbative anomalies~\cite{Klinkhamer:2003hz}.
The fermionic vacuum polarization energies of QED flux--tubes, and
abelian flux--tubes more generally, were investigated using
heat--kernel methods~\cite{Bordag:1998tg,Bordag:2003at}, world line 
numerics~\cite{Langfeld:2002vy} 
as well as the phase shift method~\cite{Graham:2004jb}.
The heat--kernel method was also used to study self--dual 
vortices~\cite{AlonsoIzquierdo:2004ru}.
This method and the world line approach limit renormalization to the 
subtraction of the divergences in the heat--kernel expansion,
thereby  obscuring the connection to perturbative renormalization. As
we will explain later, the phase shift approach is capable of making 
straightforward contact with any renormalization condition that is 
formulated in terms of (momentum space) Green's functions.
In lower dimensions the ultra--violet divergences are less severe,
which made the computation feasible for the case of two spatial 
dimensions~\cite{Graham:2006qt}, while the obstacles that arise 
for the physical case became soon obvious~\cite{Schroder:2007xk}.
A first attempt at a full calculation of the quantum 
corrections to the $Z$--string 
energy was carried out in ref.~\cite{Groves:1999ks}.
Those authors were only able to compare the energies of two
string configurations, rather than comparing a single string
configuration to the vacuum; these limitations arise from subtleties
of the renormalization process that we address in this paper.
Also, the contribution of bosonic fluctuations to the 
vacuum polarization energies have been estimated for vortex configurations
using the heat--kernel method~\cite{Bordag:2002sa} and string backgrounds
within the phase shift approach~\cite{Baacke:2008sq}, which we will use 
here for the fermionic fluctuations. Stability of cosmic string currents 
was considered in~\cite{Davis:1999ec}.

We begin by expressing the vacuum polarization energy as the renormalized 
sum over (half) the change of the single particle energies caused by
the localized background. Previously, the fermionic vacuum polarization 
energy of strings has been computed for the case of two spatial 
dimensions~\cite{Graham:2006qt}. The main purpose of the present paper 
is to extend this computation to the physical case of 
three spatial dimensions, yielding the vacuum polarization
energy per unit length of the string. Though that
extension seems straightforward since the string is translationally
invariant with respect to this additional coordinate, a number of
obstacles arise. They are mainly related to the more complex structure of
ultra--violet divergences.

It is well established that the vacuum polarization energies of 
extended background field configurations, such as solitons or vortices, 
are unambiguously obtained from a momentum integral that involves the 
derivative of the phase shifts in the potential generated by the 
background field~\cite{Graham:2009zz}. These phase shifts measure 
the distortion of the spectrum of quantum fluctuations caused by the 
background. If an object is translationally invariant with respect to a 
subset of the coordinates, we can use the phase shifts calculated in 
the nontrivial dimensions combined with appropriate kinematic coefficients 
to describe the full spectrum of 
fluctuations~\cite{Graham:2001dy}. These 
coefficients vary only with the number of trivial dimensions, but 
not with the background field. In addition to integrating the result 
over the magnitude of momentum $k$, we also need to sum over angular 
momentum channels. It has previously been shown that for string--type 
configurations, these two operations are not absolutely convergent, and 
inappropriately exchanging them may yield an unphysical 
convergence~\cite{Schroder:2007xk}. In $D=2+1$ spacetime 
dimensions, the problem is mitigated because the relevant Feynman diagram 
is manifestly finite. In $D=3+1$, the complicated structure of the 
divergent third and fourth order Feynman diagrams makes the exchange 
unavoidable. This problem is not specific to the string, and can only 
be avoided by analytic continuation to complex momenta. This procedure 
requires a careful construction of the Jost function for the Dirac 
scattering problem. On the other hand, the integration over imaginary 
momenta automatically includes the bound state contribution to the vacuum 
polarization energy~\cite{Bordag,Graham:2009zz}.

A second, more severe obstacle is specific to the string configuration. 
The method for computing the vacuum polarization energy also requires 
the evaluation of Feynman diagrams whose external legs are given in 
terms of Fourier transforms of the background field. In this calculation,
the leading terms of the Born series are subtracted and the corresponding 
contributions to the vacuum polarization energy are added back in as
(renormalized) Feynman diagrams. We stress that the energy is not computed 
from this perturbation expansion; rather, the advantage of our approach is 
that this expansion is used to make contact with standard procedures of 
perturbative renormalization, which in particular allows the implementation 
of any renormalization scheme that is formulated for the momentum space 
Green's functions of the theory. The problem that arises for the string is 
that the configuration does not vanish at spatial infinity, and so its 
Fourier transform is ill--defined. While gauge--invariant combinations 
of the Higgs and vector fields do vanish at spatial infinity, the Feynman 
series is not gauge invariant term by term. To circumvent this problem, 
we need to introduce a {\it return string}, localized at a some distance 
$\rho_0$ that is large compared to the typical extension $w$ of the 
physical string. This return string unwinds the physical string so that 
the Fourier transforms and thus the individual Feynman diagrams give 
well--defined functionals. Even though the vacuum  polarization energy 
$E_{\rm vac}$ is a non--local functional, for $\rho_0\gg w$ there should 
be a separation of scales,
\begin{equation}
E_{\rm vac}=E_{\rm ps}(w)+E_{\rm rs}(\rho_0)
\label{sepscale}
\end{equation}
for the energies of the physical string ($E_{\rm ps}$) and that of
the return string ($E_{\rm rs}$), since the overlap between the 
associated densities vanishes. Then computing $E_{\rm vac}$ and 
finding $E_{\rm rs}(\rho_0)$ for a special case, yields $E_{\rm ps}(w)$,
the quantity that we are seeking. A similar approach has been previously 
used to compute the vacuum polarization energy of electromagnetic flux 
tubes. In that case dimensional analysis indicated that 
$\lim_{\rho_0\to\infty}E_{\rm rs}(\rho_0)=0$, which was then confirmed by 
directly calculating the energy density~\cite{Graham:2004jb}. Here 
the situation is more complicated, due to the presence of the Higgs 
field. A na\"{\i}ve analysis indicates that in general 
$E_{\rm rs}(\rho_0)$ actually grows quadratically with $\rho_0$.
If the Higgs field of the return string stays on the chiral 
circle (meaning that its magnitude is fixed at its vacuum expectation 
value, so that only its isospin orientation changes in space),
$E_{\rm rs}(\rho_0)$ remains constant. However, there is no 
{\it a priori} reason for it to vanish. As an
aside we note that, in contrast to the QED case, the classical
energy of the return string does not vanish as $\rho_0\to\infty$.
So we need to find the energy of the return string, which
then is subtracted from eq.~(\ref{sepscale}) to separate the energy
of the physical string. Unfortunately, the general return string
background induces potentials in the scattering problem that behave 
like $1/\rho^2$ at small distances. Though the corresponding scattering 
matrix can still be computed, the evaluation of the Jost function 
remains obscure because the bounds needed to prove the analytic 
structure from the iterative solution are no longer satisfied. One way 
out of the dilemma would be to compute the return string energy within 
a covariant expansion that is reliable for large
$\rho_0$~\cite{Ball:1988xg}, but we have a better option. At least 
in a particular gauge (and we have to choose one anyhow to do the 
calculation) the small distance singularity disappears for return 
strings that have support only on the chiral circle. As a result we 
can calculate $E_{\rm rs}$ in this case, and numerically we
find that it vanishes as $\rho_0\to\infty$ in the $\overline{\rm MS}$  
renormalization scheme. This is not in contradiction to the general 
analysis that yields a constant, because the non--zero term is omitted 
in $\overline{\rm MS}$.

Another technical obstacle that did not emerge in the case of two 
space dimensions is the increased computational effort needed. In the 
case of three space dimensions, the third and fourth
order Feynman diagrams induce logarithmic divergences. When folded
with the Fourier transforms of the background potentials, these
diagrams become higher--dimensional integrals that are cumbersome to
treat numerically. We therefore introduce a {\it fake boson} field
whose second order Feynman diagram possesses the identical logarithmic 
divergence~\cite{Farhi:2001kh,Farhi:2003iu}. Since it is a solution
to a well--defined scattering problem, we are ensured of the
analytic properties of the  resulting Jost function and we can deal
with it on equal footing as the one from the Dirac problem.

In this paper we have collected the ingredients for computing the 
vacuum polarization energy of the string. For simplicity we use the
$\overline{\rm MS}$ renormalization scheme and leave the use of 
physical (on--shell) conditions and dynamical questions such as the
stabilization of the string by bound fermions to a forthcoming paper.

To be specific, we
consider a left--handed $SU(2)$ gauge theory in which a fermion doublet
$\Psi$ is coupled to the gauge field $W_\mu$ and the Higgs field
$\phi$.  Since we are studying only fermionic fluctuations, we can ignore
the self-interactions of the Higgs and gauge fields and simply assume
the Higgs potential is minimized at a nonzero vacuum expectation value
$|\phi|=v$.  We express the Higgs doublet $\phi = 
\begin{pmatrix} \phi_+ \cr \phi_0\end{pmatrix}$ as the matrix
\begin{equation}
\Phi=\begin{pmatrix}
\phi_0^* & \phi_+ \cr -\phi_+^* & \phi_0 \end{pmatrix} \,.
\label{higgs}
\end{equation}
We then have the interaction Lagrangian
\begin{equation}
\mathcal{L}=\overline{\Psi}i\gamma_\mu D^\mu P_L \Psi
+\overline{\Psi}i\gamma_\mu \partial^\mu P_R \Psi
-f\,\overline{\Psi}\left(\Phi P_R+\Phi^\dagger P_L\right)\Psi\,,
\label{gaugelag}
\end{equation}
where $P_{R,L}=\frac{1}{2}\left(1\pm\gamma_5\right)$ are projection
operators on right-- and left--handed components, respectively.  
The Higgs--fermion interaction is parameterized by the Yukawa 
coupling constant $f$, while the gauge coupling constant is defined
via the covariant derivative
\begin{equation}
D^\mu=\partial^\mu-igW^\mu \,.
\label{covder}
\end{equation}
Note that the vector fields are $2\times2$ matrices in (weak) isospace.
To simplify the model, we have taken the Weinberg angle to be zero, so
that there is no photon coupling, and assumed that the fermion doublet is
degenerate in mass.

Finally we list the counterterms that are required to compute
the vacuum polarization energy,
\begin{equation}
\mathcal{L}_{\rm ct}=
c_1{\rm tr}_I \left[W_{\mu\nu}W^{\mu\nu}\right]
+\frac{c_2}{2}\,{\rm tr}_I\left[\left(D_\mu\Phi\right)^\dagger
\left(D^\mu\Phi\right)\right]
+\frac{c_3}{2}\,{\rm tr}_I\left[\Phi^\dagger \Phi-v^2\right]
+\frac{c_4}{4}\left({\rm tr}_I\left[\Phi^\dagger \Phi-v^2\right]\right)^2\,,
\label{lct}
\end{equation}
where the gauge field tensor is $W_{\mu\nu}=\partial_\mu
W_\nu-\partial_\nu W_\mu -ig\left[W_\mu,W_\nu\right]$ for the vector
field. As indicated, the traces are with respect to isospin. Of course,
the counterterms are just the typical gauge--invariant
terms appearing in the classical Lagrangian for the Higgs and gauge
fields.  Note also that the $c_3$ and $c_4$ counterterms cancel
quadratic ultra--violet divergences, while $c_1$ and $c_2$
cancel logarithmic divergences. In a model with only 
two spatial dimensions, $c_1$ and $c_2$ would be finite, though
not necessarily zero \cite{Graham:2006qt}.

The remainder of the paper is organized as follows. In the next section 
we introduce the string configuration and discuss the obstacles mentioned 
above in more detail. In section III we briefly explain the phase 
shift method for computing the vacuum polarization energy. In particular 
we discuss the interface method for the case when the background 
configuration is translationally invariant in one coordinate. We also 
introduce the fake boson technique for simplifying the higher--order 
divergences. In section IV we explain the use of the return string on 
the chiral circle. Numerical results are presented in section V, 
and concluding remarks and an outlook for applications  are given in 
section VI.

\section{The Z--String}

The string configuration only depends on the distance $\rho$ from 
the symmetry axis (which we choose to be $\hat{z}$) and the corresponding 
azimuthal angle $\varphi$. It is characterized by a non--vanishing angular 
dependence at spatial infinity. In matrix notation we write in temporal 
gauge ($W^0=0$)
\begin{eqnarray}
\vec{W}&=&n\,{\rm sin}(\xi_1)\frac{f_G(\rho)}{\rho}\hat{\varphi}
\begin{pmatrix}
{\rm sin}(\xi_1) & i {\rm cos}(\xi_1)\,{\rm e}^{-in\varphi} \cr
-i {\rm cos}(\xi_1)\,{\rm e}^{in\varphi} & - {\rm sin}(\xi_1)
\end{pmatrix}
\qquad {\rm and} \quad \cr\cr
\Phi&=&vf_H(\rho)
\begin{pmatrix}
{\rm sin}(\xi_1)\, {\rm e}^{-in\varphi} & -i {\rm cos}(\xi_1) \cr
-i {\rm cos}(\xi_1) & {\rm sin}(\xi_1)\, {\rm e}^{in\varphi} 
\end{pmatrix}\,.
\label{string}
\end{eqnarray}
This configuration is commonly called a Z--string, because the corresponding
component $Z\propto {\rm tr}_I(W\tau_3)$ exhibits the spatial dependence
of an abelian string. The radial functions $f_G(\rho)$ and $f_H(\rho)$ 
approach unity at spatial infinity while they vanish at the origin ($\rho=0$). 
They are the typical profiles of the Nielson--Oleson 
string~\cite{Nielsen:1973cs}. We parameterize these profile functions via
\begin{equation}
f_H(\rho)=1-{\rm e}^{-\frac{\rho}{w_H}}
\qquad {\rm and} \qquad
f_G(\rho)=1-{\rm e}^{-\left(\frac{\rho}{w_G}\right)^2}\,.
\label{para1}
\end{equation}
Then the fields $\vec{W}$ and $\Phi$ are 
$\mathcal{O}(\rho)$ as $\rho\to0$ and ambiguities resulting from an
ill--defined azimuthal angle do not arise. We have also introduced a 
general winding number $n$ for the string, though in the numerical 
calculation we will only  consider $n=1$. In ref.~\cite{Graham:2006qt} 
we treated the angle $0\le \xi_1\le \frac{\pi}{2}$ as a variational 
parameter of the string configuration. We emphasize that $\xi_1$ describes 
the orientation of the Higgs field on the chiral circle. It will prove to 
be very useful to introduce the return string by allowing it to be 
space dependent such that it vanishes at spatial infinity.

\subsection{String in the Dirac Equation}

The Dirac equation that arises from the Lagrangian is dealt with in two 
steps. The dependence on the $z$--coordinate in which the string is 
invariant separates in a pure phase factor $e^{-ip_z z}$. In the second 
step we are thus left with a two--dimensional problem\footnote{We utilize
the standard representation for the Dirac--matrices and perform 
a global chiral rotation 
$U=-\begin{pmatrix}0 & i \cr i & 0\end{pmatrix}P_L+P_R$ 
to restore the typical form of the Dirac equation.}
\begin{eqnarray}
H&=&-i\begin{pmatrix}0 & \vec{\sigma}\cdot\hat{\rho} \cr
\vec{\sigma}\cdot\hat{\rho} & 0\end{pmatrix} \partial_\rho
-\frac{i}{\rho}\begin{pmatrix}0 & \vec{\sigma}\cdot\hat{\varphi} \cr
\vec{\sigma}\cdot\hat{\varphi} & 0\end{pmatrix} \partial_\varphi
+\frac{ns}{2\rho}f_G
\begin{pmatrix}-\vec{\sigma}\cdot\hat{\varphi}
& \vec{\sigma}\cdot\hat{\varphi} \cr
\vec{\sigma}\cdot\hat{\varphi}
& -\vec{\sigma}\cdot\hat{\varphi}\end{pmatrix}I_G \cr\cr
&&\hspace{2cm}
+m f_H\left[c\begin{pmatrix} 1 & 0 \cr 0 &-1\end{pmatrix}
+s\begin{pmatrix}0 & 1 \cr -1 & 0\end{pmatrix}iI_P\right]\,,
\label{eqDirac}
\end{eqnarray}
where $s={\rm sin}(\xi_1)$ and $c={\rm cos}(\xi_1)$. The explicit
matrices in the above equation refer to the spinor indices and
$\sigma_i$ are the $2\times2$ Pauli matrices. Furthermore we 
have introduced the isospin matrices
\begin{equation}
I_G(\varphi)=\begin{pmatrix}-s & -ic\,{\rm e}^{in\varphi} \cr
ic\,{\rm e}^{-in\varphi} & s \end{pmatrix}_I
\qquad \mbox{and}\qquad
I_P(\varphi)=\begin{pmatrix} 0 & {\rm e}^{in\varphi} \cr
{\rm e}^{-in\varphi} & 0 \end{pmatrix}_I\,.
\label{ioperators}
\end{equation}
If $E$ is an eigenvalue of $H$, then the single particle energy is 
$\pm\sqrt{E^2+p_z^2}$. Calculating the vacuum polarization energy 
requires us to integrate over $p_z$ in the framework of the interface 
formalism, which we review below. The sign degeneracy arises from the 
anti--commutator $\left\{H,\gamma_3\right\}=0$.

It is straightforward to extract the fermion--string interaction
from eq.~(\ref{eqDirac})
\begin{equation}
H_{\rm int}=\frac{ns}{2\rho}f_G
\begin{pmatrix}-\vec{\sigma}\cdot\hat{\varphi}
& \vec{\sigma}\cdot\hat{\varphi} \cr
\vec{\sigma}\cdot\hat{\varphi}
& -\vec{\sigma}\cdot\hat{\varphi}\end{pmatrix}I_G 
+m \left[\left(cf_H-1\right)\begin{pmatrix} 1 & 0 \cr 0 &-1\end{pmatrix}
+sf_H\begin{pmatrix}0 & 1 \cr -1 & 0\end{pmatrix}iI_P\right]\,,
\label{DiracInt}
\end{equation}
In view of the asymptotic behavior of the radial functions $f_{G,H}$ 
mentioned above, many of the problems with computing the vacuum 
polarization energy become immediately obvious. The interaction
does not vanish at spatial infinity, but rather approaches a pure
gauge configuration. This implies that there is no straightforward
way to set up a Born series, whose necessity we will recognize in
the following section. Furthermore we cannot compute any Fourier
transformation of the interaction, which makes it impossible to
use Feynman diagrams to impose conventional renormalization
schemes such as on--shell or $\overline{\rm MS}$.

It is tempting to perform a (singular) gauge transformation 
\begin{equation}
\widetilde{H}=U^\dagger H U \qquad {\rm with}\qquad
U=\begin{pmatrix}s\,{\rm e}^{-in\varphi} & -ic \cr
-ic & s\,{\rm e}^{in\varphi} \end{pmatrix}_I P_L+P_R
\label{singularGT}
\end{equation}
such that
\begin{equation}
\widetilde{H}_{\rm int}=\frac{ns}{2\rho}\left(f_G-1\right)
\begin{pmatrix}-\vec{\sigma}\cdot\hat{\varphi}
& \vec{\sigma}\cdot\hat{\varphi} \cr
\vec{\sigma}\cdot\hat{\varphi}
& -\vec{\sigma}\cdot\hat{\varphi}\end{pmatrix}I_G
+m \left(f_H-1\right)\begin{pmatrix} 1 & 0 \cr 0 &-1\end{pmatrix}\,.
\label{DiracIntGT}
\end{equation}
However, this only shifts the problem to the origin. As 
$\lim_{\rho\to0}f_G=0$ the $\frac{1}{\rho}$ term induces a $\frac{1}{\rho^2}$ 
potential in the corresponding second order Schr\"odinger type 
scattering problem. This is outside the standard regime in which 
analyticity of the Jost function is established~\cite{Chadan:1977pq}.
We comment on the use of $\widetilde{H}_{\rm int}$ at the end of 
section V.

\subsection{Unwinding the String}

To set up the scattering problem we must find a way to treat  
eq.~(\ref{DiracInt}). A possible solution to that problem can 
be envisaged immediately. If the angle $\xi_1$ were not constant, 
but a radial function going from its value at the origin (which 
defines the configuration of interest) to zero at some large distance 
$\rho_0$, the scattering problem would be 
treatable without altering the magnitude of the Higgs field. This
configuration corresponds to a return string on the chiral circle that 
has less severe ultra--violet singularities, since the $c_3$ and $c_4$ 
counterterms in eq.~(\ref{lct}) vanish on the chiral circle. For 
convenience we parameterize $\sin \xi_1$ rather than $\xi_1$ itself,
\begin{equation}
s(\rho)={\rm sin}(\xi_1)\,\frac{1-
{\rm tanh}\left(w_0\,\frac{\rho-\rho_0}{\rho_0}\right)}
{1+{\rm tanh}(w_0)}
\qquad {\rm and} \qquad
c(\rho)=\sqrt{1-s^2(\rho)}\,.
\label{para2}
\end{equation}
Of course, the chiral circle condition continues to hold. This addition
does not alter the form of the Dirac Hamiltonian, eq.~(\ref{DiracInt}). 
However, it must be taken into account that the isospin matrix $I_G$ now 
has radial dependence. The combination of eqs.~(\ref{para1}) 
and~(\ref{para2}) then describes the physical string for $\rho\ll\rho_0$ 
and a non--interacting theory for $\rho\gg\rho_0$. To accomplish the 
separation of scales indicated in eq.~(\ref{sepscale}) we require 
$\rho_0\gg w_{G,H}$. With these prerequisites, the Hamiltonian, 
eq.~(\ref{eqDirac}), has a well--defined scattering problem
that we will now describe.

\subsection{Scattering off the Unwound String}

As a first step we introduce grand--spin type states that couple spin 
and isospin, to account for the angular dependence. For fixed angular 
momentum $\ell$ there are four of them,
\begin{equation}
\begin{array}{ll}
\langle \varphi;SI|\ell + +\rangle=
{\rm e}^{i(\ell+n)\varphi}\,
\begin{pmatrix}1 \cr 0 \end{pmatrix}_S
\otimes\begin{pmatrix}1 \cr 0 \end{pmatrix}_I & \qquad
\langle \varphi;SI|\ell + -\rangle =
-i\,{\rm e}^{i\ell\varphi}\,
\begin{pmatrix}1 \cr 0 \end{pmatrix}_S
\otimes\begin{pmatrix}0 \cr 1 \end{pmatrix}_I\cr\cr
\langle \varphi;SI|\ell - +\rangle =
i\,{\rm e}^{i(\ell+n+1)\varphi}\,
\begin{pmatrix} 0 \cr 1\end{pmatrix}_S
\otimes\begin{pmatrix}1 \cr 0 \end{pmatrix}_I & \qquad
\langle \varphi;SI|\ell - -\rangle =
{\rm e}^{i(\ell+1)\varphi}\,
\begin{pmatrix}0 \cr 1\end{pmatrix}_S
\otimes\begin{pmatrix}0 \cr 1 \end{pmatrix}_I\,.
\end{array}
\label{eq:GSstates}
\end{equation}
where $S$ and $I$ refer to the spin and isospin subspaces, respectively.
These grand--spin states serve to construct the Dirac spinors in 
coordinate space,
\begin{equation}
\begin{array}{ll}
\langle \rho |++\rangle =
\begin{pmatrix}f_1(\rho)|\ell + +\rangle \cr
g_1(\rho)|\ell - +\rangle \end{pmatrix} & \qquad
\langle \rho |+-\rangle =
\begin{pmatrix}f_2(\rho)|\ell + -\rangle \cr
g_2(\rho)|\ell - -\rangle \end{pmatrix}
\cr \cr
\langle \rho |-+\rangle =
\begin{pmatrix}f_3(\rho)|\ell - +\rangle \cr
g_3(\rho)|\ell + +\rangle \end{pmatrix}& \qquad
\langle \rho |--\rangle =
\begin{pmatrix}f_4(\rho)|\ell - -\rangle \cr
g_4(\rho)|\ell + -\rangle \end{pmatrix}\,,
\end{array}
\label{eq:GSspinors}
\end{equation}
where we have suppressed the angular momentum index of the radial
functions because the Dirac equation is diagonal
in this quantum number.
We combine these eight radial functions into two vectors
\begin{equation}
\vec{f}=\begin{pmatrix}
f_1(\rho) \cr f_2(\rho) \cr f_3(\rho) \cr f_4(\rho) 
\end{pmatrix}
\qquad {\rm and} \qquad
\vec{g}=\begin{pmatrix}
g_1(\rho) \cr g_2(\rho) \cr g_3(\rho) \cr g_4(\rho)
\end{pmatrix}
\label{vecnot}
\end{equation}
to write the Dirac equation as a set of eight coupled first order 
linear differential equations in the matrix form
\begin{eqnarray}
(E-m)\,\vec{f}&=&V_{uu}\,\vec{f}+\left(D_u+V_{ud}\right)\,\vec{g}\cr
(E+m)\,\vec{g}&=&\left(D_d+V_{du}\right)\,\vec{f}+V_{dd}\,\vec{g}
\label{DiracMatrix}\,.
\end{eqnarray}
The derivative operators are fully contained in the diagonal 
matrices
\begin{eqnarray}
D_u&=&{\rm diag}\left(
\partial_{\rho}+\frac{\ell+n+1}{\rho},
\partial_{\rho}+\frac{\ell+1}{\rho},
-\partial_{\rho}+\frac{\ell+n}{\rho},
-\partial_{\rho}+\frac{\ell}{\rho}\right) \cr\cr
D_d&=&{\rm diag}\left(
-\partial_{\rho}+\frac{\ell+n}{\rho},
-\partial_{\rho}+\frac{\ell}{\rho},
\partial_{\rho}+\frac{\ell+n+1}{\rho},
\partial_{\rho}+\frac{\ell+1}{\rho}\right)\,.
\label{DuDdMatrix}
\end{eqnarray}
We will give the explicit form of the real $4\times4$ matrices $V_i$ in 
terms of the radial functions when we set up the Born series for the 
scattering data. Here it suffices to note that these matrices are real 
and vanish at spatial infinity, so the asymptotic solutions are cylindrical 
Bessel and Hankel functions. In particular, the Hankel functions
\begin{eqnarray}
\mathcal{H}_u&=&\mbox{diag}\left(
H^{(1)}_{\ell+n}(k\rho),H^{(1)}_{\ell}(k\rho),
H^{(1)}_{\ell+n+1}(k\rho),H^{(1)}_{\ell+1}(k\rho)\right) \cr
\mathcal{H}_d&=&\mbox{diag}\left(
H^{(1)}_{\ell+n+1}(k\rho),H^{(1)}_{\ell+1}(k\rho),
H^{(1)}_{\ell+n}(k\rho),H^{(1)}_{\ell}(k\rho)\right)
\label{Smat2}
\end{eqnarray}
that parameterize the outgoing asymptotic fields with (radial) 
momentum $k$ can be used to set up the scattering problem via the matrix 
generalization
\begin{equation}
\vec{f}\quad \longrightarrow\quad
\mathcal{F} \cdot \mathcal{H}_u
\qquad {\rm and} \qquad
\vec{g}\quad \longrightarrow\quad
\kappa\,\mathcal{G} \cdot \mathcal{H}_d\,.
\label{Smat1}
\end{equation}
Note that $\kappa={\rm sgn}(E)\,\sqrt{\frac{E-m}{E+m}}$ is well
defined for either sign of the energy eigenvalue since $|E|>m$ for the
scattering solution. Using the dispersion relation for real momenta 
$E^2=k^2+m^2$, we may also write $\kappa=\frac{k}{E+m}=\frac{E-m}{k}$.
The boundary conditions for the $4\times4$ complex
matrices $\mathcal{F}$ and $\mathcal{G}$ are simply 
$\lim_{\rho\to\infty} \mathcal{F} =\ID$ and 
$\lim_{\rho\to\infty} \mathcal{G} =\ID$ so the various columns of the 
above products refer to outgoing waves in different
channels. Demanding finally that the scattering wave--functions
are regular at the origin defines the scattering matrix. It can
be obtained from either the upper or lower components
\begin{eqnarray}
\mathcal{S}&=&-\lim_{\rho\to0}\,
\mathcal{H}_u^{-1}\cdot\mathcal{F}^{-1}\cdot
\mathcal{F}^*\cdot\mathcal{H}_u^\ast \cr
\mathcal{S}&=&-\lim_{\rho\to0}\,
\mathcal{H}_d^{-1}\cdot\mathcal{G}^{-1}\cdot
\mathcal{G}^*\cdot\mathcal{H}_d^\ast \,.
\label{Smat3}
\end{eqnarray}
This construction ensures that the physical scattering 
solution,\footnote{Since the matrices $V_i$ are real, 
$\mathcal{F}^*\cdot\mathcal{H}_u^\ast$ and
$\mathcal{G}^*\cdot\mathcal{H}_d^\ast$ also solve the Dirac equation.}
\begin{equation}
\Psi_{\rm s.c.}^{(u)}=\mathcal{F}^*\cdot\mathcal{H}_u^\ast
+\mathcal{G}\cdot\mathcal{H}_u\cdot\mathcal{S}\,,\qquad
\Psi_{\rm s.c.}^{(d)}=\mathcal{G}^*\cdot\mathcal{H}_u^\ast
+\mathcal{F}\cdot\mathcal{H}_u\cdot\mathcal{S}\,.
\label{sctsol}
\end{equation}
is regular at the origin, $\rho\to0$.
An important numerical check (besides unitarity) is that both equations
yield identical scattering matrices. We write the Dirac equation for
the matrices $\mathcal{F}$ and $\mathcal{G}$ in a form that will later 
simplify the Born series,
\begin{eqnarray}
\partial_\rho \mathcal{F} &=&
\left[\overline{\mathcal{M}}_{ff}+O_d\right]\cdot\mathcal{F}
+\mathcal{F}\cdot\mathcal{M}_{ff}^{(r)}
+\left[\overline{\mathcal{M}}_{fg}+kC\right]\cdot\mathcal{G}\cdot Z_d\,\cr
\partial_\rho \mathcal{G} &=&
\left[\overline{\mathcal{M}}_{gg}+O_u\right]\cdot\mathcal{G}
+\mathcal{G}\cdot\mathcal{M}_{gg}^{(r)}
+\left[\overline{\mathcal{M}}_{gf}-kC\right]\cdot\mathcal{F}\cdot Z_u\,.
\label{deqforGF}
\end{eqnarray}
The matrices 
\begin{eqnarray}
C&=&{\rm diag}(-1,-1,1,1)\,,\cr \cr
O_u&=&\fract{1}{\rho}\,{\rm diag}\,
\left(-(\ell+n+1),-(\ell+1),\ell+n,\ell\right)\,,\cr\cr
O_d&=&\fract{1}{\rho}\,{\rm diag}\,
\left(\ell+n,\ell,-(\ell+n+1),-(\ell+1)\right)\,,\cr\cr
Z_d&=&{\rm diag}\, \left(
\frac{H^{(1)}_{\ell+n+1}(k\rho)}{H^{(1)}_{\ell+n}(k\rho)}\,,
\frac{H^{(1)}_{\ell+1}(k\rho)}{H^{(1)}_{\ell}(k\rho)}\,,
\frac{H^{(1)}_{\ell+n}(k\rho)}{H^{(1)}_{\ell+n+1}(k\rho)}\,,
\frac{H^{(1)}_{\ell}(k\rho)}{H^{(1)}_{\ell+1}(k\rho)}\right)
=\left(Z_u\right)^{-1}\,,\cr\cr
\mathcal{M}_{gg}^{(r)}&=&kC\cdot Z_u-O_u\ \quad {\rm and} \quad
\mathcal{M}_{ff}^{(r)}=-kC\cdot Z_d-O_d\,,
\label{kinmatrix}
\end{eqnarray}
are purely kinematic, while the background fields are contained in
the real--valued $4\times4$ matrices
\begin{eqnarray}
\overline{\mathcal{M}}_{gg}&=&\begin{pmatrix}
\alpha_G R_+ & \alpha_P R_- \cr
-\alpha_P R_- & -\alpha_G R_+ \end{pmatrix} \qquad
\overline{\mathcal{M}}_{gf}=\frac{1}{\kappa}\left[
\alpha_HC+ \alpha_G\begin{pmatrix} 0 & -R_+ \cr
R_+ & 0 \end{pmatrix}\right]\qquad \cr \cr
\overline{\mathcal{M}}_{ff}&=&
\begin{pmatrix}
-\alpha_G R_+ & \alpha_P R_- \cr
-\alpha_P R_- & \alpha_G R_+ \end{pmatrix} \qquad\quad
\overline{\mathcal{M}}_{fg}=
\kappa\left[
\alpha_HC- \alpha_G\begin{pmatrix} 0 & -R_+ \cr
R_+ & 0 \end{pmatrix}\right] \,,
\label{bornmatrices}
\end{eqnarray}
where
\begin{equation}
R_+=\begin{pmatrix} s & c \cr c & -s \end{pmatrix} \qquad
R_-=\begin{pmatrix} 0 & -1 \cr 1 & 0 \end{pmatrix} \,.
\label{defrpm}
\end{equation}
The profile functions enter via
\begin{equation}
\alpha_H=m\left(cf_H-1\right)\,,\quad \alpha_P=msf_H
\quad \mbox{and}\quad
\alpha_G=\frac{ns}{2\rho}\,f_G\,.
\label{radfunct}
\end{equation}
Note that the inclusion of the return string makes $R_+$ a function 
of $\rho$, while $R_-$ remains constant in space.

The Born series is a straightforward but tedious expansion in the 
matrices $\overline{\mathcal{M}}_i$, which are linear in the 
background fields. For example, at first order we have to solve
\begin{eqnarray}
\partial_\rho \mathcal{F}_1 &=&
O_d\cdot\mathcal{F}_1+\mathcal{F}_1\cdot\mathcal{M}_{ff}^{(r)}
+kC\cdot \mathcal{G}_1\cdot Z_d+\overline{\mathcal{M}}_{ff}
+\overline{\mathcal{M}}_{fg}\cdot Z_d\cr
\partial_\rho \mathcal{G}_1 &=&
O_u\cdot\mathcal{G}_1+\mathcal{G}_1\cdot\mathcal{M}_{gg}^{(r)}
-kC\cdot \mathcal{F}_1\cdot Z_u+\overline{\mathcal{M}}_{gg}
+\overline{\mathcal{M}}_{gf}\cdot Z_u\,,
\label{firstborn}
\end{eqnarray}
with the asymptotic conditions $\lim_{\rho\to\infty}\mathcal{F}_1=0$
and $\lim_{\rho\to\infty}\mathcal{G}_1=0$. The first order correction to 
the scattering matrix then becomes
$\mathcal{S}_1=\lim_{\rho\to0} \mathcal{H}_u^{-1}\left[
\mathcal{F}_1-\mathcal{F}_1^\ast\right]\mathcal{H}_u^\ast$.
The solutions
$\mathcal{F}_1$ and $\mathcal{G}_1$ are the sources for the second
order terms $\mathcal{F}_2$ and $\mathcal{G}_2$, yielding the 
second Born approximation 
$\mathcal{S}_2=\lim_{\rho\to0} \mathcal{H}_u^{-1}\left[
\mathcal{F}_1\left(\mathcal{F}_1^\ast-\mathcal{F}_1\right)
+\mathcal{F}_2-\mathcal{F}_2^\ast\right]\mathcal{H}_u^\ast$.
This iteration in the background fields can be repeated to any
desired order, yielding the Born series for the scattering matrix
$\mathcal{S}=\ID+\mathcal{S}_1+\mathcal{S}_2+\ldots\,$.

\subsection{Jost Function}

Ultimately we need to compute the phase of ${\rm det}(\mathcal{S})$, 
which according to eq.~(\ref{Smat3}) is related to the phase of 
$F=F_\ell(k)={\rm det}\left(\lim_{\rho\to0}\mathcal{F}\right)$, or 
equivalently $G=G_\ell(k)={\rm det}(\lim_{\rho\to0}\mathcal{G})$. 
Furthermore, we need to sum over angular momentum $\ell$ and integrate 
over radial momentum $k$ after subtracting sufficiently many terms 
of the Born series. This procedure is not only numerically cumbersome 
because these are oscillating functions of $k$, but it may even 
cause erroneous results~\cite{Schroder:2007xk} since these 
sums/integrals are not absolutely convergent. These obstacles are avoided 
by analytically continuing to imaginary momenta $t=ik$ and performing
the integrals along the branch cut $t>m$. The analytic continuation
for the Dirac equation is conceptually different from the well--studied
Schr\"odinger case because $E=\pm\sqrt{k^2+m^2}$ causes the complex 
momentum plane to have two sheets. So on the real axis we have to pick 
one sign, continue to complex momenta and compute the Jost function on 
the imaginary axis. This procedure must then be repeated for the other 
sign and then all discontinuities must be collected at the end. In the 
present problem we are fortunate because the solutions to the Dirac 
equation exhibit charge conjugation symmetry along the real axis. 
Therefore ${\rm det}(\mathcal{S})$ does not change under $E\to-E$ and 
there is no additional discontinuity in the Jost function. Moreover, 
the Jost function is real on the imaginary axis, as in the Schr\"odinger 
problem. However, the way this comes about in the string problem 
requires us to be careful when constructing the Jost function for 
complex momenta.

We describe the case with $E=\sqrt{k^2+m^2}$. The analytic 
continuation concerns the Hankel functions, which turn
into modified Bessel functions: $Z_u\to Y_u$ and $Z_d\to Y_d$, with
\begin{equation}
Y_u={\rm diag}\, \left(
\frac{K_{\ell+n}(t\rho)}{K_{\ell+n+1}(t\rho)}\,,
\frac{K_{\ell}(t\rho)}{K_{\ell+1}(t\rho)}\,,
-\frac{K_{\ell+n+1}(t\rho)}{K_{\ell+n}(t\rho)}\,,
-\frac{K_{\ell+1}(t\rho)}{K_{\ell}(t\rho)}\right) =
-\left(Y_d\right)^{-1}\,.
\label{defZ12}
\end{equation}
Furthermore we have the change of the kinematic coefficient 
$\kappa\to z_\kappa$, with
\begin{equation}
z_\kappa=\frac{m+i\sqrt{t^2-m^2}}{t} \,,
\label{kappaAC}
\end{equation}
a pure phase. The differential equations, which we solve
subject to the boundary conditions that $\mathcal{F}$ and
$\mathcal{G}$ approach unity at spatial infinity, read
\begin{eqnarray}
\partial_\rho \mathcal{F} &=&
\left[\overline{\mathcal{M}}_{ff}+O_d\right]\cdot\mathcal{F}
+\mathcal{F}\cdot\mathcal{M}_{ff}^{(r)}
+\left[\overline{\mathcal{M}}_{fg}-tC\right]\cdot\mathcal{G}\cdot Y_d\cr
\partial_\rho \mathcal{G} &=&
\left[\overline{\mathcal{M}}_{gg}+O_u\right]\cdot\mathcal{G}
+\mathcal{G}\cdot\mathcal{M}_{gg}^{(r)}
+\left[\overline{\mathcal{M}}_{gf}+tC\right]\cdot\mathcal{F}\cdot Y_u\,.
\label{deqforGFAC}
\end{eqnarray}
For simplicity we have omitted the momentum arguments in the radial 
wave--functions $\mathcal{F}$ and $\mathcal{G}$. It should be apparent
from the context that these are distinct from the solutions to 
eqs.~(\ref{deqforGF}). The coefficient matrices are slightly modified:
\begin{equation}
\begin{array}{ll}
\mathcal{M}_{gg}^{(r)}=-tC\cdot Y_u-O_u & \qquad
\mathcal{M}_{ff}^{(r)}=tC\cdot Y_d-O_d \cr
\overline{\mathcal{M}}_{gf}=z_\kappa\left[
\alpha_HC+ \alpha_G\begin{pmatrix} 0 & -R_+ \cr
R_+ & 0 \end{pmatrix}\right] & \qquad
\overline{\mathcal{M}}_{fg}=
-z_\kappa^*\left[
\alpha_HC- \alpha_G\begin{pmatrix} 0 & -R_+ \cr
R_+ & 0 \end{pmatrix}\right]  \,,
\end{array}
\label{bornmatricesAC}
\end{equation}
while $\overline{\mathcal{M}}_{gg}$ and $\overline{\mathcal{M}}_{ff}$
are as on the real axis. Note, that in contrast to the Schr\"odinger
problem, the differential equations do not become real on the imaginary
axis. Rather, charge conjugation $E\to-E$ induces complex conjugation. It 
is therefore not surprising that the na\"{\i}ve continuation to
$\lim_{\rho\to0}\mathcal{F}$ is not real. Instead, we find numerically 
that $F=G^*$. Interestingly enough, for a given value of $t$, the imaginary 
part is the same for all angular momenta. The origin for this imaginary 
part lies in the subtle definition of the Jost function via 
the Wronskian between the Jost solution, {\it i.e.} $\mathcal{F}$
or $\mathcal{G}$, and the regular solution that satisfies
momentum independent boundary conditions at the origin. The latter are
required to ensure the regular solution to be an analytic function. 
Analyticity of the Jost solution is guaranteed by the non--singular 
behavior of the interaction potentials in the presence of the return 
string. At the origin, the Higgs field differs from its vacuum expectation 
value (it actually vanishes). In the Dirac case this modifies the relative 
weight of the upper and lower components. On the real axis this weight
is given by $\kappa$ in eq.~(\ref{Smat1}). More precisely, at the origin 
the non--diagonal elements of the matrices in eq.~(\ref{bornmatricesAC}) 
vanish and the eight differential equations decouple with respect to the 
index on the radial functions in eq.~(\ref{vecnot}), {\it i.e.}
spin and weak isospin. For real momenta a typical solution in the
vicinity of $\rho=0$ then looks like~\cite{Bordag:2003at}
\begin{equation}
\begin{pmatrix} f_4 \cr g_4 \end{pmatrix}
\sim \left(\frac{k}{q}\right)^{l}
\begin{pmatrix} \sqrt{E+mc f_H(0)}\, J_l(q\rho) \cr\cr
\sqrt{E-m cf_H(0)}\, J_{l+1}(q\rho) \end{pmatrix}
\label{regsol}
\end{equation}
with $q=\sqrt{E^2-(m c f_H(0))^2}$
and similar dependencies for the other six radial functions.
The square--root coefficients cause the proper definition of the 
Jost function, $\nu(t)$, to be
\begin{equation}
{\rm exp}\left[\nu(t)\right]
=\left(\frac{\tau-im}{\tau-im c f_H(0)}\right)^2 \lim_{\rho\to0}
{\rm det}(\mathcal{F})
=\left(\frac{\tau+im}{\tau+im c f_H(0)}\right)^2 \lim_{\rho\to0}
{\rm det}(\mathcal{G})
\label{defJost}
\end{equation}
with $\tau=\sqrt{t^2-m^2}$. The power of two occurs 
because we compute the determinant of a $4\times4$ matrix. Note that 
this redefinition not only cancels the imaginary parts, but
also modifies the real part. Furthermore
it cancels the logarithmic singularity in ${\rm ln}\lim_{\rho\to0}
{\rm det}(\mathcal{F})$ observed numerically at $t\sim m$. Since $f_H$ 
is part of the interaction, this correction factor also undergoes 
expansion in the framework of the Born series. Otherwise, {\it i.e.}
for ${\rm det}(\mathcal{F})$, the Born series is constructed as for
real momenta by iterating the differential eq.~(\ref{deqforGFAC})
in $\overline{\mathcal{M}}_i$.

The resulting Jost function is a continuous function in the upper 
complex momentum plane and the branch cuts in the Dirac equation do 
not carry over to $\nu$. This is a consequence of charge conjugation
invariance in the present model.

\section{Phase shift approach in D=3+1}

In this section we collect all the pieces needed to compute the vacuum 
polarization energy for the configuration in which the physical and 
the return strings are combined.

\subsection{Interface Formalism}

The interface formalism addresses the problem of computing the 
vacuum polarization energy for configurations that are translationally 
invariant in a subset of coordinates. As laid out in 
ref.~\cite{Graham:2001dy}, this approach only requires scattering 
data from the lower dimensional space in which the configuration is 
non--trivial. However, we still have to adapt the formalism to the present
string problem and, in particular, motivate the above recipe for 
computing the logarithmic Jost function $\nu(t)$ on the imaginary 
momentum axis. 

Our starting point is
the interface formula for the vacuum polarization energy per 
unit length of a configuration that is translationally invariant
in one coordinate,\footnote{In comparison with eq.(6) of 
ref.~\cite{Graham:2001dy} a factor two emerged from the two signs 
of the single particle energies.}
\begin{eqnarray}
\Delta E_\delta^{(N)}&=&\frac{1}{4\pi}\sum_\ell \Bigg\{D_\ell 
\int_0^\infty \frac{dk}{\pi}
\left[(k^2+m^2) {\rm ln}\left(\frac{k^2+m^2}{\mu^2}\right)-k^2\right]\,
\frac{d}{dk}\left[\delta_\ell(k)\right]_N \cr
&& \hspace{2cm}
+\sum_j\left[\left(\epsilon_{j,\ell}\right)^2
{\rm ln}\frac{\left(\epsilon_{j,\ell}\right)^2}{\mu^2}
-\left(\epsilon_{j,\ell}\right)^2+m^2\right]\Bigg\}\,.
\label{interface1}
\end{eqnarray}
A few remarks on the notation are in order. The most important input is 
the phase shift $\delta_\ell={\rm ln}{\rm det}(\mathcal{S})$ of the 
scattering matrix in eq.~(\ref{Smat3}). Using the factorization property
of the determinant and the fact that $Y_\nu(z)\gg J_\nu(z)$ for
$z\to0$ we may write
\begin{equation}
\delta_\ell(k)=\frac{1}{i}{\rm ln}\, {\rm det}
\lim_{\rho\to0}\mathcal{F}_\ell(\rho,k)^{-1}\mathcal{F}^\ast_\ell(\rho,k)
=\frac{1}{i}{\rm ln}\, {\rm det}
\lim_{\rho\to0}\mathcal{G}_\ell(\rho,k)^{-1}
\mathcal{G}^\ast_\ell(\rho,k)\,.
\label{delta}
\end{equation}
The square bracket with index $N$ in eq.~(\ref{interface1}) indicates 
that the first $N$ terms of the Born series have been subtracted. This 
is important both for the integral to converge and also to ensure
that the semi--circle at infinity does not contribute when performing
the integral by contour integration in the complex momentum plane.
Later we will add back this contribution in terms of renormalized
Feynman diagrams. This is indeed the heart of the phase shift approach, 
as it allows the implementation of standard (perturbative) renormalization
conditions. In case of the string background the degeneracy factor 
$D_\ell$ in eq.~(\ref{interface1}) is
\begin{equation}
D_\ell=\begin{cases}
1\,, \quad & \ell=-n \cr
2\,, \quad & {\rm otherwise}\,,
\end{cases}
\label{degeneracy}
\end{equation}
where $n=1$ is the winding number of the scalar field introduced in the
string background eq.~(\ref{string}).
Finally $\epsilon_{j,\ell}$ denote the bound state energies in the 
$\ell^{\rm th}$ partial wave with $\epsilon_{j,\ell}<m$.
The renormalization scale $\mu$ has no effect as a consequence of 
sum rules for scattering data~\cite{Graham:2001iv}. We nevertheless
require it for dimensional reasons and for simplicity take $\mu=m$. 

To set up the contour integral we first remark that [{\it cf.}
eq.~(\ref{kinmatrix})] $Z_d(-k)=-Z_d^\ast(k)$. This implies 
that\footnote{This also determines  
$\kappa=\frac{k}{E+m}=\frac{E-m}{k}$ as the genuine starting
equation for the analytic continuation in $k$.}
$\mathcal{F}_\ell(\rho,-k)=\mathcal{F}_\ell^\ast(\rho,k)$ since
the boundary condition for $\mathcal{F}$ at $\rho\to\infty$ is 
independent from $k$. (Identical relations hold, of course, for
$\mathcal{G}$.) We therefore write for the momentum
integral in eq.~(\ref{interface1}) (with $E=\sqrt{k^2+m^2}$)
\begin{eqnarray}
&&\int_0^\infty \frac{dk}{\pi}
\left[E^2 {\rm ln}\left(\frac{E^2}{m^2}\right)-k^2\right]\,
\frac{d}{dk}\left[\delta_\ell(k)\right]_N=\cr\cr
&&\hspace{1cm}
-\int_{-\infty}^\infty \frac{dk}{2\pi i}
\left[E^2 {\rm ln}\left(\frac{E^2}{m^2}\right)-k^2\right]\,
\frac{d}{dk}\left[{\rm ln}\,{\rm det}\mathcal{F}_\ell(0,k)\right]_N=\cr\cr
&&\hspace{1cm}
-\int_{-\infty}^\infty \frac{dk}{2\pi i}
\left[E^2 {\rm ln}\left(\frac{E^2}{m^2}\right)-k^2\right]\,
\frac{d}{dk}\left[{\rm ln}\left(\frac{E+m}{E+m c f_H(0)}\right)^2
+{\rm ln}\,{\rm det}\mathcal{F}_\ell(0,k)\right]_N
\label{interface2}
\end{eqnarray}
where the argument $\rho=0$ is understood as the limit $\rho\to0$. The 
additional term introduced in the second equation adds a piece to the 
integrand that is odd in $k$ and therfore does not change the integral. 
This modification has introduced the Jost function, since
$\nu(t)$ is precisely the analytic continuation of the quantity in
square brackets. We have already argued that $\nu(t)$ is an analytic
function in the upper half momentum plane. Hence when closing the
contour, we only need to consider the discontinuity of the logarithm
along the cut $t\ge m$, which is $2\pi i$. In addition we need to add
the contributions from poles to the contour integral. As in the
non--relativistic case,  the existence of any bound state is related
to a simple root of the Jost function at $k=i\kappa_{j,\ell}
=i\sqrt{m^2-\left(\epsilon_{j,\ell}\right)^2}$. Therefore
\begin{equation}
{\rm tr}\left[\mathcal{F}^{-1}\frac{d}{dk} \mathcal{F}\right]
\sim \frac{1}{k-i\kappa_{j,\ell}}
\label{Fpole}
\end{equation}
in the vicinity of $k=i\kappa_{j,\ell}$. The contribution of the 
resulting pole cancels against the explicit bound state contributions
in eq.~(\ref{interface1}). (This is a general property of the phase shift 
approach.) As a result, eq.~(\ref{interface1}) can be recast to
\begin{equation}
\Delta E_\delta^{(N)} = - \frac{1}{2 \pi} \int\limits_{m}^\infty
dt\,t \sum_{\ell} D_\ell \left[\nu_\ell(t)\right]_N\,.
\label{MASTER0}
\end{equation}
Here we have integrated by parts and also used the important property that
the integral along the imaginary axis can be interchanged with the angular
momentum sum~\cite{Schroder:2007xk}. A final change of variables 
$t\to\tau=\sqrt{t^2-m^2}$ yields
\begin{equation}
\Delta E_\delta^{(N)}=-\frac{1}{2\pi} \int_0^\infty d\tau\, \tau
\sum_\ell D_\ell \left[\nu_\ell(\sqrt{\tau^2+m^2})\right]_N \,.
\label{MASTER}
\end{equation}
Equation~(\ref{MASTER}) is our master formula for the vacuum polarization 
energy of the string. However, we still have to make sense of the $N$ 
subtractions.

\subsection{Feynman Diagrams in $\overline{\rm\bf MS}$}

Having subtracted the first $N$ terms of the Born series to the
vacuum polarization energy, we must add them back in form of
Feynman diagrams. We then combine these diagrams with the counterterms,
eq.~(\ref{lct}), to renormalize the theory.

We generate the Feynman diagrams by expanding the effective action
\begin{equation}
\mathcal{A}=\mathcal{A}_0-i\,{\rm Tr}{\rm ln}\left\{
1+\left(i\dslash-m\right)^{-1} H_I\right\}
\label{effa1}
\end{equation}
with the interaction
\begin{equation}
H_I=L_\mu\gamma^\mu P_L +h +ip\gamma_5
\label{hint1}
\end{equation}
and the isospin operators
\begin{equation}
\begin{array}{llll}
L_0=0\,,\quad
& \vec{L}=2\alpha_G \hat{\varphi} I_G(\varphi)\,,\quad
& h=-\alpha_H\ID\,,\quad
& p= -\alpha_P I_P(\varphi) \,.
\end{array}
\label{hint2}
\end{equation}
The isospin matrices $I_G$ and $I_P$ are defined in eq.~(\ref{ioperators}).
The first two orders of the expansion contain quadratic and 
sub--dominant logarithmic ultra--violet divergences. The third and
fourth order terms are merely logarithmically divergent. The latter
divergence can be extracted by omitting all external momenta in
comparison with the loop momentum. These are the local contributions
to the Feynman diagrams. To completely capture the divergence structure
we therefore need to compute the first and second order completely (to 
control the quadratic divergences), but we may restrict ourselves to 
the local contribution in the case of the third and fourth order terms. 
This simplifies the computation considerably. In dimensional regularization 
the divergent contributions to the action from all four orders are 
(where $L\cdot L=L_\mu L^\mu$, etc.\@)
\begin{eqnarray}
\mathcal{A}^{\rm (div)}&=&i \int d^4x
\left[\left(\frac{\mu}{m}\right)^{4-D} \int\frac{d^Dl}{(2\pi)^D}
\left(l^2-1+i\epsilon\right)^{-2}\right]
\nonumber\\*
&&\times {\rm tr}_I\Bigg\{\frac{1}{6}\left(\partial_\alpha L_\beta
-i\left[L_\alpha,L_\beta\right]\right)^2
-\left[\left(h-m\right)^2+p^2\right]L\cdot L \cr
&&\hspace{1.2cm}
-\left(\partial h\right)^2-\left(\partial p\right)^2
+2\left(h-m\right)L\cdot \partial p -iL\cdot\left[\partial p,p\right]\cr
&&\hspace{1.2cm}
+2m^2\left[\left(h-m\right)^2+p^2-m^2\right]
+\left[\left(h-m\right)^2+p^2-m^2\right]^2
\Bigg\}\,.
\label{efflocal}
\end{eqnarray}
The counterterms in eq.~(\ref{lct}) are identical to the
integrand of the spatial integral.\footnote{For example,
${\rm tr}_I \left(D_\mu \Phi\right)^\dagger 
\left(D^\mu \Phi\right)=
\frac{v^2}{m^2}\,{\rm tr}_I\big\{
\left(\partial h\right)^2+\left(\partial p\right)^2
+ \left[\left(h-m\right)^2+p^2\right] L\cdot L
-2\left(h-m\right)L\cdot \partial p +i
L\cdot\left[\partial p,p\right]\big\}$.} Hence these
divergences are removed by writing the counterterm 
coefficient as ($s=1,\ldots,4$) 
\begin{equation}
c_s=-i\left(\frac{\mu}{m}\right)^{4-D}
\int \frac{d^Dl}{(2\pi)^D}\left(l^2-1+i\epsilon\right)^{-2}
+\overline{c}_s\,,
\label{finitecoef}
\end{equation}
where the $\overline{c}_s$ are the finite parts that are fixed by the
renormalization conditions.
 
The remaining ultra--violet finite contribution from the first
and second order is
\begin{eqnarray}
\mathcal{A}^{\rm (fin)}
&=&\frac{1}{(4\pi)^2}\int d^4x\, {\rm tr}_I\big\{4m^3h\big\}
-\frac{1}{8\pi^2} \int\frac{d^4k}{(2\pi)^4} \,
\left(m^2-\frac{k^2}{6}\right) {\rm tr}_I
\left[h(k)h(-k)+p(k)p(-k)\right]\cr
&&-\frac{1}{8\pi^2} \int\frac{d^4k}{(2\pi)^4} \int_0^1 dx \,
{\rm ln}\left[1-x(1-x)\frac{k^2}{m^2}\right] \nonumber \\*
&& \hspace{0.2cm}\times {\rm tr}_I
\Bigg\{\hspace{-0.2cm}\left[m^2-x(1-x)k^2\right]
\left[\fract{1}{2} L(k)\cdot L(-k)
-3\left(h(k)h(-k)+p(k)p(-k)\right)\right] \cr
&&\hspace{1.5cm} +x(1-x)\left[k\cdot L(k)\, k\cdot L(-k)
-\fract{1}{2}k^2 L(k)\cdot L(-k)\right] \cr
&&\hspace{1.5cm}+2m^2 p(k)p(-k)
+im k\cdot L(k) p(-k) \Bigg\} \,,
\label{eff2}
\end{eqnarray}
where the Fourier transforms have been indicated by the arguments 
of the profile functions. Observe that the $m^3$ and $m^2$ terms 
in the first two integrals combine to be proportional to an integral 
over ${\rm tr}_I\left[\Phi^\dagger\Phi -v^2\right]$. Hence they cancel
exactly against the $c_3$--type counterterm if we set its finite part 
to $\overline{c}_3=\frac{1}{4\pi^2}\frac{m^4}{v^2}=
\frac{f^2m^2}{4\pi^2}$. This defines the no--tadpole renormalization
scheme, as the vacuum polarization energy is at least quadratic in
the external fields. For the remaining counterterms, we discard 
all finite pieces,
\begin{equation}
\overline{c}_1=\overline{c}_2=\overline{c}_4=0\,.
\label{msbar}
\end{equation}
This defines the $\overline{\rm MS}$ renormalization scheme
that we impose further on. From eq.~(\ref{eff2}) we then
determine the Feynman diagram contribution to the vacuum polarization
energy up to quadratic order in the profiles to be
\begin{eqnarray}
\Delta E_{\rm FD}&=&\int_0^\infty \frac{k dk}{4\pi} \,\Bigg\{
\frac{k^2}{3}\left(h_0^2+p_n^2\right)
+4m^2I_1p_n^2+2mkI_1\alpha_c^{(+)}p_n
+k^2I_2\left[\left(\alpha_c^{(+)}\right)^2
\hspace{-0.1cm}-\left(\alpha_c^{(-)}\right)^2
\hspace{-0.1cm}-\left(\alpha_s\right)^2\right]
\cr
&&\hspace{2cm}-\left(m^2I_1+k^2I_2\right)
\left[6h_0^2+6p_n^2+\left(\alpha_c^{(+)}\right)^2
+\left(\alpha_c^{(-)}\right)^2+\left(\alpha_s\right)^2\right]
\Bigg\}\,.
\label{fdint0}
\end{eqnarray}
with the parameter integrals ($\eta=k/m$)
\begin{eqnarray}
I_1&=&\int_0^1 dx \, {\rm ln}\left[1+x(1-x)\eta^2\right]
=\frac{2}{\eta}\sqrt{4+\eta^2}\,
{\rm arsinh}\left(\fract{\eta}{2}\right)-2\,,\cr\cr
I_2&=&\int_0^1 dx \, x(1-x)\,{\rm ln}\left[1+x(1-x)\eta^2\right]
=\frac{\sqrt{4+\eta^2}}{3\eta^3}\left[\eta^2-2\right]
{\rm arsinh}\left(\fract{\eta}{2}\right)+\frac{2}{3\eta^2}-\frac{5}{18}\,,
\label{fdint1}
\end{eqnarray}
and the Fourier transforms that contain the background profiles,
\begin{eqnarray}
h_0(k)&=&\int_0^\infty \rho\, d\rho\, \alpha_H(\rho)J_0(k\rho)\,, \qquad
p_n(k)=\int_0^\infty \rho\, d\rho\, \alpha_P(\rho)J_n(k\rho)\,,\cr
\alpha_c^{(\pm)}(k)&=&\int_0^\infty \rho\, d\rho\,\alpha_G(\rho)c(\rho)
\left[J_{n-1}(k\rho)\pm J_{n+1}(k\rho)\right]\,,\cr
\alpha_s(k)&=&\int_0^\infty \rho\, d\rho\,\alpha_G(\rho)s(\rho)
J_1(k\rho)\,.
\label{ft}
\end{eqnarray}
We still have to work out the third and fourth order pieces, for 
which we want to avoid computing the full Feynman diagrams. Essentially
we only need to compensate the logarithmic divergence 
\begin{equation}
\mathcal{A}^{\rm (div)}_{3,4}= \pi  c_F\, TL\,
\left[i\left(\frac{\mu}{m}\right)^{4-D}
\int\frac{d^Dl}{(2\pi)^D}\left(l^2-1+i\epsilon\right)^{-2}\right]
\label{logdiv}
\end{equation}
where $T$ and $L$ are the (infinite) lengths of the time and 
$z$--axis intervals. The constant of proportionality is 
a simple integral over the profile functions,
\begin{eqnarray}
c_F&=&\int_0^\infty \rho\, d\rho\,\Big\{
\left(\alpha_H^2+\alpha_P^2\right)
\left(\alpha_H^2+\alpha_P^2+4m\alpha_H\right)
+4\left(\alpha_H^2+\alpha_P^2+2m\alpha_H\right)\alpha_G^2
\nonumber \\*
&&\hspace{3cm}-\frac{4n}{\rho}\,\alpha_G\alpha_P
\left(s\alpha_P+c\alpha_H\right)\Big\}\,.
\label{limcoef}
\end{eqnarray}
This result implies the limit, {\it cf.} eq.~(\ref{MASTER0}),
\begin{equation}
\sum_{\ell}D_\ell \,t\, [\nu_\ell(t)]_2\,
\stackrel{t\to\infty}{\mbox{\Large $\longrightarrow$}}\, -\frac{c_F}{t}\,,
\label{limfct}
\end{equation}
which is well suited to verify the accuracy of the numerical results.

\subsection{Fake Boson}

As already mentioned, dimensional analysis of the Dirac problem
suggests that $N=4$ subtractions are necessary in
eqs.~(\ref{interface1}) and~(\ref{MASTER}). This
is unfortunate since the Feynman diagrams for $N=3$ and $N=4$ are
complicated higher dimensional integrals (in Fourier space and over Feynman
parameters). Also the corresponding Born orders are numerically costly
when integrating the Dirac equation. As discussed above, we would like 
to employ an alternative method to deal with the logarithmic
ultra--violet divergences that emerge at these orders. Here we will
briefly explain one possibility. An important prerequisite for this, 
however, is the above analysis on the imaginary axis that allows us 
to perform the angular momentum sum {\it prior} to the momentum 
integral~\cite{Schroder:2007xk}.

Essentially we do not want to subtract the third and fourth
order terms of the Born series in eq.~(\ref{MASTER}), but
some other quantity that satisfies the following three conditions,
\begin{itemize}
\item[1)] it must exhibit the same analytic properties as
the $N=3,4$ terms of the Born series,
\item[2)] it must cancel the logarithmic divergence at large $t$: 
asymptotically it must behave like 
$\frac{c_F}{t}$,
\item[3)] we must be able to add it back as a sum of Feynman diagrams.
\end{itemize}
The perfect candidate that automatically satisfies conditions 1) and 3) is 
the second order contribution to the vacuum polarization energy of a 
background potential $V$ coupled to a fluctuating boson field. To ensure 
the second condition, we merely need to fix the strength of the
background potential such that the divergence of the corresponding
Feynman diagram matches eq.~(\ref{logdiv}).  For definiteness, we take 
an exponential potential
\begin{equation}
V(\rho)=m^2\frac{\rho}{\rho_0}\,
{\rm e}^{-2\rho/\rho_0}\,.
\label{fakepot}
\end{equation}
that also is translationally invariant along the $z$--axis. We choose
the scale to be set by the return string because it determines the
regions of momenta and angular momenta that dominate the integral
and sum in eq.~(\ref{MASTER}). We call $\overline{\nu}_\ell(t)$
the Jost function of this boson problem on the imaginary axis 
and $\overline{\nu}_\ell^{(2)}(t)$ its second Born approximation 
and define
\begin{equation}
\nu(t)=\lim_{\ell_{\rm max}\to\infty}\sum_{\ell=-n}^{\ell_{\rm max}}
D_\ell\left[\nu_\ell(t)\right]_2
+\frac{c_F}{c_B}\,
\lim_{\overline{\ell}_{\rm max}\to\infty}
\sum_{\ell=0}^{\overline{\ell}_{\rm max}}
\left(2-\delta_{\ell,0}\right)\overline{\nu}_\ell^{(2)}(t)\,.
\label{finalJost}
\end{equation}
Here 
\begin{equation}
c_B=\frac{1}{4}\int_0^\infty \rho\, d\rho\, V^2(\rho)
=\frac{3m^4\rho_0^2}{512}
\label{boslogdiv}
\end{equation}
is the boson analogue to eq.~(\ref{limcoef}). We have taken care to 
allow the (numerical) cut--offs for the angular momentum sums to be 
different in the fermion and boson cases. We merely have to ensure 
that, at a given $t$, either sum has converged to a sufficient accuracy.

Now we can perform the momentum integral and compute
\begin{equation}
\Delta E_\delta=-\frac{1}{2\pi} 
\int_0^\infty d\tau\, \tau \nu(\sqrt{\tau^2+m^2})
\label{MASTER2}
\end{equation}
as the phase shift contribution to the vacuum polarization energy. 

Finally, we find the total vacuum polarization energy of the 
unwound string to be the sum of three terms
\begin{equation}
\Delta E =\Delta E_\delta+\Delta E_{\rm FD}+\Delta E_{\rm B}\,.
\label{VERYMASTER}
\end{equation}
The last term originates from the finite part of the second order
boson contribution,
\begin{equation}
\Delta E_{\rm B}=-\frac{c_F}{c_B}
\int_0^\infty \frac{k dk}{16\pi} \, I_1 V_0^2\,,
\label{fdbos}
\end{equation}
with the Fourier transform
\begin{eqnarray}
V_0(k)&=&\int_0^\infty \rho\, d\rho\, V(\rho)J_0(k\rho)=
m^2\rho_0^2\,\frac{8-k^2\rho_0^2}
{\left[4+k^2\rho_0^2\right]^{\frac{5}{2}}}\,.
\label{ftbos}
\end{eqnarray}
Two remarks on the fake boson method are in order. First, it can be 
used with any renormalization prescription because the counterterms 
are determined from the two leading orders as a consequence of 
gauge invariance. Second, it is important to have identical masses 
for fake boson loop integral as in the coefficient of 
$\overline{\nu}^2_\ell$ in eq.~(\ref{VERYMASTER}). We note 
that this fake boson procedure has already been tested in similar 
models, including a consistent treatment of its mass 
parameter~\cite{Farhi:2001kh}. 

\section{Return String in the Singular Gauge}

The techniques introduced so far allow us to compute the combined
vacuum polarization energy of the physical and return strings, but we 
still have to disentangle them. In general this is a tedious
or even impossible task. However, for the case that the return 
string is restricted to the chiral circle, the gauge--transformed
Hamiltonian, eq.~(\ref{DiracIntGT}), suggests a potentially
successful procedure. The pure return string is characterized by
$f_G=f_H=1$ which apparently leads to $\widetilde{H}_{\rm int}=0$.
This would actually imply that the return string has zero vacuum 
polarization energy. However, this conclusion is premature
since eq.~(\ref{DiracIntGT}) has been derived for a constant
angle $\xi_1$. Repeating the same gauge transformation but allowing
a radial dependence as in eq.~(\ref{para2}) yields
\begin{equation}
\widetilde{H}=-i\begin{pmatrix}0 & \vec{\sigma}\cdot\hat{\rho} \cr
\vec{\sigma}\cdot\hat{\rho} & 0\end{pmatrix} \partial_\rho
-\frac{i}{\rho}\begin{pmatrix}0 & \vec{\sigma}\cdot\hat{\varphi} \cr
\vec{\sigma}\cdot\hat{\varphi} & 0\end{pmatrix} \partial_\varphi
+m\beta +\frac{s^\prime}{2c}
\begin{pmatrix}-\vec{\sigma}\cdot\hat{\rho}
& \vec{\sigma}\cdot\hat{\rho} \cr
\vec{\sigma}\cdot\hat{\rho}
& -\vec{\sigma}\cdot\hat{\rho}\end{pmatrix}I_P\,.
\label{retsringH}
\end{equation}
Here
\begin{equation}
s^\prime=\frac{ds(\rho)}{d\rho}=
\frac{{\rm sin}(\xi_1)}{1+{\rm tanh}(w_0)}\,
\frac{w_0}{\rho_0}\,\left[
{\rm tanh}^2\left(w_0\,\frac{\rho-\rho_0}{\rho_0}\right)-1\right]\,.
\label{retinteract}
\end{equation}
As a consequence of chiral symmetry, the gauge--transformed Hamiltonian 
only has an induced vector field while the (pseudo)scalar fields have 
disappeared.

In contrast to the gauge--transformed Hamiltonian of the physical
string, eq.~(\ref{DiracIntGT}), no singularity emerges as $\rho\to0$ 
and we have a well--defined scattering problem. It can be treated in 
complete analogy to what we described in previous sections using
\begin{eqnarray}
\overline{\mathcal{M}}_{gg}&=&\alpha_r
\begin{pmatrix} R_- & 0 \cr 0 & R_-\end{pmatrix} \qquad
\overline{\mathcal{M}}_{gf}=-z_\kappa\alpha_r
\begin{pmatrix} 0 & R_- \cr  R_- &0 \end{pmatrix}  \cr \cr
\overline{\mathcal{M}}_{ff}&=&\alpha_r
\begin{pmatrix} R_- & 0 \cr 0 & R_-\end{pmatrix} \qquad
\overline{\mathcal{M}}_{fg}=z_\kappa^\ast\alpha_r
\begin{pmatrix} 0 & R_- \cr  R_- & 0 \end{pmatrix}\,,
\label{bornmatricesreturn}
\end{eqnarray}
where $\alpha_r=\frac{s^\prime}{2c}$ sets the order parameter of the 
corresponding Born series. Note that these are the coefficient 
matrices on for imaginary momenta, {\it i.e.} they are to be used
in eq.~(\ref{deqforGFAC}).

The induced vector field only has a single isospin orientation, so
the commutator term in the corresponding field tensor vanishes and only 
second order pieces contribute in eq.~(\ref{efflocal}). As a result, the
ultra--violet divergences from the third and fourth order Born terms
vanish, which simplifies the renormalization procedure for the return 
string considerably. In particular, we do not need to apply the fake 
boson procedure.

For the return string we do not need to take special measures to 
impose the no--tadpole condition because the $c_3$--type counterterm 
vanishes on the chiral circle. Hence in the $\overline{\rm MS}$ scheme
the vacuum polarization energy of the return string becomes
\begin{equation}
\Delta E^{\rm (r.s.)}=
-\frac{1}{2\pi}
\int_0^\infty d\tau\, \tau 
\sum_{\ell}D_\ell\left[\widetilde{\nu}_\ell(\sqrt{\tau^2+m^2})\right]_2
+\Delta E^{\rm (r.s.)}_{\rm FD}\,,
\label{returnenergy}
\end{equation}
where $\widetilde{\nu}_\ell$ is the Jost function associated 
with the Hamiltonian, eq.~(\ref{retsringH}), and 
\begin{equation}
\Delta E^{\rm (r.s.)}_{\rm FD}=
-\int \frac{kdk}{4\pi}
\Big\{m^2I_1\left[\left(\alpha_r^{(+)}\right)^2+
\left(\alpha_r^{(-)}\right)^2\right]
+2k^2I_2\left(\alpha_r^{(-)}\right)^2\Big\}
\label{returnFD}
\end{equation}
is the finite piece of the second order Feynman diagram.
It includes the Fourier transforms
\begin{equation}
\alpha_r^{(\pm)}(k)=\int_0^\infty \rho d\rho\,\alpha_r(\rho)
\left[J_{n+1}(k\rho)\pm J_{n-1}(k\rho)\right]\,.
\label{retFT}
\end{equation}
We are now in a position to present numerical results.

\section{Numerical Analysis}

In this section we present results for the vacuum polarization energy 
\begin{equation}
E_{\rm ps}=\Delta E - \Delta E^{\rm (r.s.)}
\label{efinal}
\end{equation}
of the non--abelian string generated by the fluctuating fermions.
The fermion mass $m=vf$ sets the scale for all data presented in this
section. The length parameters $w_{G,H}$ and $\rho_0$ are measured
in units of $\frac{1}{m}$, and since $E_{\rm ps}$ is the vacuum 
polarization energy per unit length of the string, it is measured in $m^2$.
Note that the inverse width parameter ($w_0$) associated with the 
return string is dimensionless. As mentioned earlier, all results 
presented here are for unit winding of the string, $n=1$.
 
The main purpose of the present study is to demonstrate, as a 
proof--of--principle, 
that the leading (fermion) contribution to the vacuum
polarization energy of a string in a non-abelian gauge theory
can indeed be unambiguously computed. This has made necessary the
introduction of a return string. We parameterized its shape by
two variables, the position $\rho_0$ and the (inverse) width $w_0$. 
In this study we concentrate on the variation of 
$E_{\rm ps}$ on these two parameters, to establish a well--defined 
limit. The underlying feature of our calculations is the 
separation of the physical and return strings' vacuum energies.
Before presenting the corresponding results, we like to discuss
a necessary condition for this separation to be observed. The physical
string generates a zero mode for $\xi_1=\frac{\pi}{2}$ in the 
$\ell=-n$ channel~\cite{Naculich:1995cb}, regardless of the specific 
shape that is parameterized by $w_H$ and $w_G$. This zero mode must 
also be produced when the physical string is augmented by the return 
string if the separation hypothesis holds, because the bound state 
wave function does not interfere with the return string. In
table~\ref{tab_bs} we present the lowest single particle 
energy\footnote{We compute this energy by discretizing the 
spectrum of the free Dirac Hamiltonian in a large cylindrical cavity,
computing the interaction matrix elements from those eigenstates,
and diagonalizing the resulting matrix.} for $\xi_1=\frac{\pi}{2}$ as 
a function of the separation between the physical and the return string.
\begin{table}[t]
\begin{tabular}{c|ccccc}
$\rho_0$ & 6 & 8 & 10 & 12 & 14 \cr
\hline
$\epsilon_0$ & 0.0065 & 0.0024 & 0.0014 & 0.0010 & 0.0008 
\end{tabular}
\caption{\label{tab_bs}\sf Lowest bound state energy for 
$\xi_1=\frac{\pi}{2}$, $w_H=w_G=2$ and $w_0=8$ as a function of
the position of the return string, $\rho_0$.}
\end{table}
As expected, this eigenvalue approaches zero as the separation
increases. These results provide confidence in the separation hypothesis
already at moderate values of $\rho_0$.

The numerical calculation of the vacuum polarization energy is 
quite involved. The main effort concerns the phase shift part, 
$\Delta E_\delta$ in eq.~(\ref{VERYMASTER}). In addition to the 
criteria mentioned after eqs.~(\ref{Smat3}) and~(\ref{limfct}) we 
have furthermore tested these calculations of scattering data with 
respect to
\begin{itemize}
\item charge conjugation symmetry: $E\to-E$ (for real momenta)
\item reflection around the winding of the string: $\ell \to -\ell-n$
\item sum rules for distinct angular momenta, {\it i.e.} agreement
of integrals involving the Jost--function over real and
imaginary momenta.
\end{itemize}

We integrate the differential eqs.~(\ref{deqforGF}),
(\ref{deqforGFAC}), their Born expansions as well as the fake boson 
and return string analogues from some large radius 
$\rho_{\rm max}\sim 4\rho_0$ to $\rho_{\rm min}\sim0$ with the 
boundary condition $\mathcal{F}(\rho_{\rm max},k)=\ID$, and
identify $\lim_{\rho\to0}\mathcal{F}(\rho,k)
=\mathcal{F}(\rho_{\rm min},k)$. Alternatively, this identification
can equally well be obtained from the derivative of the 
wave--function. Furthermore, a differential equation is formulated
for ${\rm ln}\,{\rm det} \mathcal{F}(\rho,k)^{-1}
\mathcal{F}(\rho,k)^\ast$ to avoid $2\pi$
ambiguities in the computation of the phase shift, $\delta_\ell(k)$,
{\it cf.} eq~(\ref{delta}). The computations for real momenta have
been performed mainly for use in the consistency tests mentioned above.
Channels that include (modified) Hankel functions with zero index
($\ell=-2,-1,0$) are particularly cumbersome because in separating 
the regular and irregular solutions, we need to distinguish 
${\rm ln}(\rho)$ from a constant at very small distances 
$\rho_{\rm min}$, {\it cf.} eq.~(\ref{sctsol}).
That is, $\rho_{\rm min}$ must be taken tiny to obtain
the correct scattering matrix in eq.~(\ref{Smat3}). On the
real axis, the result can be checked against extracting the 
$\mathcal{S}$--matrix from the derivative of the scattering 
wave--function. For calculations on the imaginary axis we
assume $\rho_{\rm min}\sim 10^{-60}$ and successively carry out
an extrapolation
\begin{equation}
\nu(\rho_{\rm min})=\nu_0 +\frac{a_1}{{\rm ln}(\rho_{\rm min})}
+\frac{a_2}{{\rm ln}^2(\rho_{\rm min})}\ldots\,\,,
\label{nuextra}
\end{equation}
for the Jost function in these channels. We test the final result,
{\it i.e.} $\nu_0$, for stability against further changes of 
$\rho_{\rm min}$ and also check the condition ${\sf Im}(\nu_0)=0$. 
In other channels $\rho_{\rm min}\sim 10^{-12}$ suffices to represent 
the origin.

For the sum over angular momentum channels, we must go to very large 
channel numbers (typically several hundred). The required value 
$\ell_{\rm max}$ to obtain convergence increases with $\rho_0$. While 
this is expected, since $\rho_0$ determines the impact parameter, 
it limits our ability to send $\ell_{\rm max}$ to infinity. We 
therefore perform an extrapolation to $\ell_{\rm max}\to\infty$. 
Using various analytical forms for this extrapolation we estimate 
an absolute error of $0.005$ for $\Delta E_\delta$. Fortunately, the 
right hand side of eq.~(\ref{finalJost}) approaches zero at moderate 
momenta, so that we can approximate the upper limit of the integral in 
eq.~(\ref{MASTER2}) by $10$ or even less.

The individual sums in eq.~(\ref{finalJost}) have typical maximal 
values of the order of tens, while they combine to a result at the 
order of a tenth or less. This behavior is typical for
these type of computations because the physics information is hidden
beneath the ultra--violet divergences. This disproportion increases
with the distance of the return string. For practical purposes,
this unfortunately implies a severe loss of precision of
three orders of magnitude or more. However, demanding too high
numerical precision makes the integration of the Dirac equation
(and its Born expansion) ineffective. Fortunately, doing the
momentum integral on the imaginary axis regains quite some
efficency because we are not plagued by oscillating integrands.

In total, it takes several CPU--days on a modern PC to 
generate a number for $\Delta E$ for a single set of 
parameters ($w_{G,H},\xi_1,w_0,\rho_0$). The fake boson piece
$\sum_\ell \overline{\nu}_\ell^2$, in comparison, is cheap since
it takes less than a CPU--hour even though 
$\overline{\ell}_{\rm max}\approx 3\ell_{\rm max}$ for 
sufficient convergence.

We first consider the return string contribution, which only depends 
on the parameters $w_0$ and $\rho_0$. The results are shown in 
table~\ref{tab_rs}.
\begin{table}
\centerline{
\begin{tabular}{cc|cc|c}
$w_0$ & $\rho_0$ & $\Delta E^{\rm (r.s.)}_{\delta}$ 
& $\Delta E^{\rm (r.s.)}_{\rm FD}$ & $\Delta E^{\rm (r.s.)}$\cr
\hline
8 & 6  & -0.006 & -0.100 & -0.107 \cr
8 & 8  & -0.004 & -0.048 & -0.052 \cr
8 & 10 & -0.003 & -0.027 & -0.030 \cr
8 & 12 & -0.002 & -0.018 & -0.020 \cr
8 & 14 & -0.002 & -0.012 & -0.014 \cr
\hline
6 & 6  & -0.003 & -0.037 & -0.040 \cr
6 & 8  & -0.002 & -0.018 & -0.020 \cr
6 & 10 & -0.001 & -0.011 & -0.013 \cr
6 & 12 & -0.001 & -0.007 & -0.008 \cr
6 & 14 & -0.001 & -0.005 & -0.006
\end{tabular}}
\caption{\label{tab_rs}\sf Numerical results for the
vacuum polarization energy of the return string in the
$\overline{\rm MS}$ scheme. The entry $\Delta E^{\rm (r.s.)}_{\delta}$
refers to the $\tau$ integral in eq.~(\ref{returnenergy}). It is the 
analogue of eq.~(\ref{MASTER2}) for the sole return string.
Scales are set by the fermion mass $m$.}
\end{table}
We observe that $\Delta E^{\rm(r.s.)}$ is small in magnitude and quickly 
decreases as the position of the return string is sent to infinity.
Dimensional analysis indicates that except for the Higgs kinetic term 
(the $c_2$--type in eq.~(\ref{lct})) all contributions from the return 
string vanish as $\rho_0\to\infty$. This is the local part of the second
order Feynman diagram, which is taken out in the
$\overline{\rm MS}$ scheme. Hence both the smallness and the
large $\rho_0$ behavior of $\Delta E^{\rm(r.s.)}$ are well
understood. Since for our particular parameterization, the
relevant parameter is $w_0/\rho_0$, the above argument
also explains the decrease with $w_0$ that can be observed from
table~\ref{tab_rs}. In essence, the contribution of the sole return
string is well under control.

In table~\ref{tab1_comb} we show the contributions to the vacuum 
polarization energy of the configuration consisting of both the physical 
and the return string, for various parameters that describe the return 
string part of the configuration. We observe a significant cancellation 
between the integral over the subtracted Jost--function and the fermion 
Feynman diagram contribution. This behavior, of course, can be different 
in renormalization schemes other than $\overline{\rm MS}$. The fake boson 
Feynman diagram contribution is negligible in the $\overline{\rm MS}$ scheme.
\begin{table}
\centerline{
\begin{tabular}{cc|ccc|c}
$w_0$ & $\rho_0$ & $\Delta E_\delta$ & $\Delta E_{\rm FD}$ &
$\Delta E_{\rm B}$ & $\Delta E$ \cr
\hline
8 & 6  & 0.168 & -0.120 & 0.013 & 0.061 \cr
8 & 8  & 0.153 & -0.105 & 0.012 & 0.060 \cr
8 & 10 & 0.144 & -0.096 & 0.013 & 0.061 \cr
8 & 12 & 0.132 & -0.090 & 0.013 & 0.055 \cr
8 & 14 & 0.121 & -0.085 & 0.013 & 0.049 \cr
\hline
6 & 6  & 0.148 & -0.080 & 0.012 & 0.080 \cr
6 & 8  & 0.137 & -0.072 & 0.012 & 0.077 \cr
6 & 10 & 0.128 & -0.067 & 0.011 & 0.072 \cr
6 & 12 & 0.118 & -0.064 & 0.012 & 0.066 \cr
6 & 14 & 0.107 & -0.061 & 0.012 & 0.057 \cr
\end{tabular}}
\caption{\label{tab1_comb}\sf Numerical results for the vacuum 
polarization energy of the combination of the physical and the
return strings in the $\overline{\rm MS}$ scheme with
$w_H=w_G=2$ and $\xi_1=0.4\pi$.}
\end{table}

In table~\ref{tab_ps1} we combine the results from the previous
two tables to find the vacuum polarization energy of the physical
string. Certainly the data indicate a saturation for large
$\rho_0$, as demanded by the assumption that the scales for the 
physical and return strings separate
\begin{table}
\centerline{
\begin{tabular}{cc|cc|c}
$w_0$ & $\rho_0$ & $\Delta E$ & $-\Delta E^{\rm (r.s.)}$
& $\Delta E_{\rm ps}$\cr
\hline
8 & 6  & 0.061 & 0.107 & 0.168 \cr
8 & 8  & 0.060 & 0.052 & 0.112 \cr
8 & 10 & 0.061 & 0.030 & 0.091 \cr
8 & 12 & 0.055 & 0.020 & 0.075  \cr
8 & 14 & 0.049 & 0.013 & 0.062  \cr
\hline
6 & 6  & 0.080  & 0.040 & 0.120 \cr
6 & 8  & 0.077  & 0.020 & 0.097 \cr
6 & 10 & 0.072  & 0.014 & 0.086 \cr
6 & 12 & 0.066  & 0.008 & 0.074 \cr
6 & 14 & 0.057  & 0.006 & 0.063
\end{tabular}}
\caption{\label{tab_ps1}\sf Vacuum polarization energy of the physical string
for $w_H=2$, $w_G=2$ and $\xi_1=0.4\pi$.}
\end{table}
This separation assumption is further supported by the observation that 
for increasing distance of the return string, its width ($\frac{1}{w_0}$) 
no longer affects the energy of the physical string. In this limit, the 
vacuum polarization energy of the string turns out to be less than a tenth 
of the fermion mass. We have repeated this analysis for a thin string and 
present the results in table~\ref{tab_ps2}.
\begin{table}
\centerline{
\begin{tabular}{c|cc|c}
$\rho_0$ & $\Delta E$ & $-\Delta E^{\rm (r.s.)}$
& $\Delta E_{\rm ps}$\cr
\hline
6  &  0.003& 0.107 & 0.110 \cr
8  &  0.022& 0.052 & 0.074 \cr
10 &  0.029& 0.030 & 0.059 \cr
12 &  0.027& 0.020 & 0.047 \cr
14 &  0.025& 0.013 & 0.038 \cr
\end{tabular}}
\caption{\label{tab_ps2}\sf Same as table~\ref{tab_ps1} for
$w_H=0.5$, $w_G=1$ and $\xi_1=0.4\pi$. Here we used $w_0=8$.}
\end{table}
We find that its vacuum polarization energy is even smaller.

We recall that for typical model parameters, {\it i.e.} gauge and Yukawa
couplings\footnote{The relevant scale is set by the heaviest fermion
coupling to the string, {\it i.e.} the top quark, where
$f\sim175/190\sim1$ as the Weinberg angle is set to zero.} of the 
order unity or less, the classical energy of the physical 
string configuration, which we did not discuss here, is
of the order $1$ or $10$. Also typical single 
particle energies of the fermions bound in the string background are 
around $0.5$. Hence the vacuum polarization energy in the 
$\overline{\rm MS}$ renormalization scale is about an order of 
magnitude smaller than other relevant energy scales in the model. 
This smallness of string (vortex) vacuum polarization energies has 
also been observed for string type configurations in simpler 
models~\cite{Bordag:2003at} and hence is not fully unexpected.

We have also performed numerical calculations with the singular 
Hamiltonian of eq.~(\ref{DiracIntGT}) by regularizing the 
singularity of the gauge coupling:
$\frac{1}{\rho}\to\frac{1}{\rho+w_{\rm reg}}$. 
In the second order scattering problem the unregularized form
induces a $\frac{1}{\rho^2}$ divergence at small $\rho$, which 
eventually spoils the analytic properties of the scattering data.
However, we did not obtain a smooth limit as we removed the regulator. 
This is not a contradiction to the existence of a well defined 
vacuum polarization energy. It just implies that two operations of
(i) integrating to $\rho\to0$ and (ii) removing the regulator
may not be exchanged. This can already be observed from computing
the classical energy with such a regulator. The spatial integral,
which involves the field strength tensor squared, has no well--defined 
limit as the regulator $w_{\rm reg}$ is sent to zero.

Finally, we would like to mention that we have also considered
the numerical computation on the real axis, {\it e.g.} using
eq.~(\ref{interface1}), to compute the vacuum polarization
energy for this Dirac problem. As expected from simpler models, we 
find that a twofold subtraction suffices to render a convergent
integral when the angular momentum sum is carried out first. This 
result contradicts the ultra--violet divergence structure of the 
quantum field theory. It is a mathematical artifact stemming from 
the improper exchange of sum and integrals~\cite{Schroder:2007xk}. 
This has been the major reason to analytically continue to imaginary 
momenta, since it allows this exchange. We remark that a rigorous Born 
subtraction, {\it i.e.} up to at least fourth order in 
$\overline{\mathcal{M}}_i$ in eq.~(\ref{deqforGF}), can indeed be 
carried out on the real momentum axis. As argued before this seems 
technically problematic. It is the exchange of sums and integrals 
that requires absolute convergence, which is only guaranteed for 
imaginary momenta.

\section{Conclusions}

We have presented a feasible method for the computation of the 
fermion contribution to the vacuum polarization energy of a string 
type configuration in a non--abelian gauge theory. If there are many 
internal degrees of freedom for the fermions, for example color, 
this contribution is the dominant quantum correction to the energy 
of the string. The key tool that we employed was the introduction of 
a {\it return string} configuration on the chiral circle. It was, in 
particular, indispensable to obtain a well--defined Feynman series, 
which is required to impose conventional renormalization conditions.
For the present study we used the $\overline{\rm MS}$ scheme, augmented
by the no--tadpole condition. The generalization to other schemes,
such as on--shell, does not cause further (technical) subtleties.
The advantage of having the return string on the chiral circle
is that its vacuum polarization energy can be reliably computed in 
a specific gauge. Then it suffices to compute the energy of any
physical string configuration with respect to this return string. 
Furthermore, technical innovations such as the computation of
momentum integrals via analytic continuation and the introduction of 
a fake boson field have been necessary to reduce the computational 
workload, making the investigation feasible with appropriate expenditure 
of computational resources.

Our results suggest that for future investigations of the total energy 
of string configurations it is a good approximation to omit
the $\overline{\rm MS}$ contribution to the vacuum polarization
energy. This contribution is about an order of magnitude smaller than
other typical energies in the model. This is in particular the case
for a thin string. Such an approximation simplifies the computation of
the energy of the string tremendously: It merely remains to find the
(finite pieces of) the counterterm coefficients, since the spatial 
integral of these terms is, by the definition of renormalizability,
the same as for the classical energy. The remaining spatial integrals 
only involve gauge invariant combinations and thus can immediately
be computed {\it without} the return string. This result will make
the search for configurations feasible that are stabilized on
purely dynamical grounds. For instance, such a stabilization can 
occur by trapping fermions along the string. Once such a configuration 
is discovered, it remains to verify that its vacuum polarization 
energy in the $\overline{\rm MS}$ scheme indeed is tiny, which can 
be achieved with the techniques presented here. We leave this study 
to a forth--coming paper.

\section*{Acknowledgments}
We thank M. Bordag for interesting discussions.
This work was supported by the National Science Foundation (NSF)
through grants PHY05-55338 and PHY08-55426 (NG).

\end{document}